\documentclass[12pt]{article}
\usepackage{amssymb}
\usepackage{epsfig,color}
\usepackage{pstricks,graphicx,epsfig,color,amssymb,amsmath,amscd}
\usepackage{cite}
\usepackage[]{graphicx}
\usepackage{makeidx}
\usepackage{amsmath}
\usepackage{nccmath}
\usepackage{setspace}
\usepackage{ulem}
\usepackage{tensor}
\usepackage{bm}
\usepackage{indentfirst}
\usepackage{sectsty}
\subsectionfont{\normalfont\itshape\bfseries}

\usepackage[]{caption}
\renewcommand\rho{\varrho}
\newcommand{\bi}{\bibitem}
\newcommand{\be}{\begin{eqnarray}}
\newcommand{\ee}{\end{eqnarray}}

\renewcommand{\thesection}{\arabic{section}.}

\begin{document}

\begin{titlepage}
\date{}
\title{Calculations of scalaron decay probabilities}
\author{E.\,V. Arbuzova$^{1, 2}$, A.\,D. Dolgov$^{2, 3}$, A.\,S. Rudenko$^{2, 4,}$\footnote{E-mail: a.s.rudenko@inp.nsk.su}}

\maketitle
\begin{center}
$^1${Department of Higher Mathematics, Dubna State University,\\ 
Universitetskaya st.\,19, Dubna, 141983 Russia}\\
$^2${Department of Physics, Novosibirsk State University,\\ 
Pirogova st.\,2, Novosibirsk, 630090 Russia}\\
$^3${Bogoliubov Laboratory of Theoretical Physics, Joint Institute for Nuclear Research,\\
Joliot-Curie st.\,6, Dubna, Moscow region, 141980 Russia}\\
$^4${Theory Division, Budker Institute of Nuclear Physics,\\
akademika Lavrentieva prospect 11, Novosibirsk, 630090 Russia} 
\end{center}

\vspace{5mm}

\textbf{Abstract}---The particle production through the scalaron decays is considered for several different channels. The central part of the work is dedicated to a study of the decay probability into two complex minimally coupled massless scalars. The calculations are performed by two different independent methods. In addition we calculated the decay probability into real minimally coupled massless scalars, conformally coupled massive scalars, massive fermions, and gauge bosons. The results are compared with the published papers which in some cases disagree with each other.

\thispagestyle{empty}

\end{titlepage}

\section{Introduction}

\indent The popular now mechanism of the Starobinsky inflation is based on the introduction of the additional term quadratic in scalar curvature, $R$, into the canonical Hilbert--Einstein action~\cite{AAS-R2}:
\be
S (R^2) = -\frac{M_\text{Pl}^2}{16\pi} \int d^4 x \sqrt{-g}\,\left[R- \frac{R^2}{6M_R^2}\right],
\label{action-R2}
\ee
where $M_\text{Pl} \approx 1.2 \times 10^{19}$ GeV is the Planck mass and $M_R$ is a constant parameter with dimension of mass. According to the estimate of~\cite{faulkner} the magnitude of temperature fluctuations of cosmic microwave background (CMB) demands $M_R \approx 3 \times 10^{13} $ GeV. 

Nonlinear in curvature terms (as well as the terms containing Ricci and Riemann tensors) arise as a result of radiation corrections to the energy--momentum tensor of matter in curved space--time~\cite{ginzburg} (see also~\cite{AD-YZ}, section XVI). However such terms appear naturally with the normalizing mass of the order of the Planck mass, $M_\text{Pl} \sim 10^{19}$ GeV, while the necessary value of $M_R$ in action~\eqref{action-R2} is about 5--6 orders of magnitude lower.

Due to the nonlinear term in the action curvature, $R$ becomes a dynamical variable and we can speak about new gravitational scalar degree of freedom, scalaron, with mass equal to $M_R$.

As is argued e.g. in review~\cite{ADR-symmetry}, cosmological evolution in $R^2$-modified theory is naturally divided into the following four epochs: \\
1) inflation, when $R$ slowly decreases from some large value $R/M_R^2 \gtrsim 10^2$, \\
2) curvature oscillations, which at $M_R t \gg 1$ (here and below time $t = 0$ corresponds to the onset of oscillation phase) are described by the expression:
\be
R(t) = - 4 M_R\, \frac{\cos (M_R t + \theta)}{t},
\label{R-of-t}
\ee
leading to efficient particle production through the scalaron decay and consequently to the universe heating; the Hubble parameter at this stage behaves as~{\cite{ADS-distorted} }
\be
H \equiv \frac{\dot a}{a} = \frac{2}{3t}\, \left[ 1 + \sin \left(M_R t + \theta\right) \right],
\label{H-of-t} 
\ee 
3) transition of the scalaron domination regime to the dominance of the produced matter of mostly relativistic particles,
and \\
4) transition to the conventional cosmology governed by the General Relativity. 

In this paper we confine ourselves to the epoch of the universe heating and calculate the rate of the production of different types of particles with the aim to resolve some discrepancies in the existing literature. The main attention is paid to the case of scalaron decays into complex minimally coupled massless scalars, which has not been previously considered in the literature. We perform the calculations in two different independent ways: the usual calculations of the matrix element of the external field $R(t)$ between vacuum and a pair of the scalar particle state and calculating quantum corrections to the scalaron equation of motion. The latter method is analogous to those considered in~\cite{Dolgov_Hansen,AD-Freese,EA-AD-LR}. We have also studied scalaron decays into real minimally coupled massless scalars, conformally coupled massive scalars, massive fermions, and massless gauge bosons. In the latter case the decay is induced by the conformal anomaly. 


The action \eqref{action-R2} is presented in the so-called Jordan frame. We prefer to use it because the equation of motion of the scalaron field in this frame has the form of the usual Klein--Gordon equation, see below Eq.~\eqref{D2-R-1}. However in several papers as e.g. in~\cite{gorb-pan} the so-called Einstein frame is used. Both frames are presumably equivalent, but the equation of motion in the Einstein frame is considerably more complicated.

In the course of this paper we assume that the metric is the spatially-flat Friedmann--Lema\^itre--Robertson--Walker (FLRW) one with the interval
\be 
ds^2 = dt^2 - a^2(t) \delta_{ij} dx^i dx^j,
\label{FLRW}
\ee 
where $a(t)$ is the cosmological scale factor and the Hubble parameter is expressed through $a(t)$ as $H = \dot a/a$.

As one can see from Eq. \eqref{FLRW} the metric tensor $g_{\mu\nu}$ is taken with the signature convention $(+,-,-,-)$. The Riemann tensor describing the curvature of space--time is determined according to 
$\tensor{R}{^\alpha_{\mu\beta\nu}} = \partial_\beta \Gamma^\alpha_{\mu\nu} + \dots$, 
$R_{\mu\nu} = \tensor{R}{^\alpha_{\mu\alpha\nu}}$, and $R = g^{\mu\nu}R_{\mu\nu}$. 

Equation of motion for the curvature scalar which follows from action \eqref{action-R2} has the form:
\be
D^2 R + M_R^2 R = - \frac{8 \pi M_R^2}{M_\text{Pl}^2} \, T^\mu _\mu,
\label{D2-R-1}
\ee
where $D^2 = g^{\mu\nu} D_\mu D_\nu $, $D_\mu$ is the covariant derivative in metric \eqref{FLRW} and $T^\mu_\mu$ is the trace of the energy--momentum tensor of matter, which comes from the canonical matter action omitted in Eq.~\eqref{action-R2}. The concrete forms of the matter action are presented in what follows.
For homogeneous $R = R(t)$:
\be
D^2 R = \left(\partial_t^2 + 3 H \partial_t \right)R. 
\label{D2-R}
\ee

The effective action of the scalaron field leading to equation of motion \eqref{D2-R-1} can be taken as 
\be 
A_R = \frac{M_\text{Pl}^2}{48 \pi M_R^4}\,\int d^4x \sqrt{-g} 
\left[\frac{(DR)^2}{2} - \frac{M_R^2 R^2}{2} - \frac{ 8\pi M_R^2}{M_\text{Pl}^2}\, T^\mu_\mu R \right].
\label{A-R}
\ee

To determine the energy density of the scalaron field we have to redefine this field in such a way that the kinetic term of the new field enters the action with the usual coefficient 1/2. So the canonically normalized scalar field is~\cite{ADS-distorted}:
\be
\Phi = \frac{M_\text{Pl}}{\sqrt{48 \pi}\, M_R^2}\,R.
\label{Phi}
\ee

Correspondingly, the energy density of the scalaron field is equal to
\be
\rho_R = \rho_\Phi = \frac{\dot\Phi^2 +{M_R^2} \Phi^2}{2} = 
\frac{M_\text{Pl}^2 (\dot R^2 + {M_R^2} R^2)}{96 \pi M_R^4}.
\label{rho-R-1}
\ee

\section{Scalaron decays into scalar particles}

We assume that the actions of the non-interacting, except for coupling to gravity, complex and real scalar fields with mass $m$ have respectively the forms: 
\be
 S_c [\phi_c]& = & \int d^4x\,\sqrt{-g}\,\left(g^{\mu\nu}\partial_\mu\phi_c^*\,\partial_\nu\phi_c - m^2 |\phi_c|^2 + \xi R |\phi_c |^2\right), 
\label{A-phi-compl}  \\
S_r [\phi_r] & = & \frac{1}{2}\int d^4x\,\sqrt{-g}\,\left(g^{\mu\nu}\partial_\mu\phi_r\,\partial_\nu\phi _r - m^2 \phi_r^2 + \xi R \phi^2_r\right).
\label{A-phi-real}
\ee
If the constant $\xi$ is zero, fields $\phi$'s are called minimally coupled to gravity; for $\xi = 1/6$ they are called conformally coupled, because in this case the traces of the energy--momentum tensors of the fields $\phi_{c,r}$ vanish.

The equation of motion both for real and complex fields $\phi$'s has the form 
\be
D^2 \phi + m^2 \phi - \xi R \phi = 0,
\label{D2-phi}
\ee
which in metric \eqref{FLRW} transforms to
\be
\ddot\phi - \frac{\Delta\phi}{a^2} + 3H\dot\phi + m^{2}\phi - \xi R\,\phi = 0, 
\label{ddot-phi}
\ee
where $\Delta$ is the three-dimensional Laplace operator in flat 3D-space.

The energy--momentum tensor of $\phi$ is defined as the variation of the action over the metric tensor: 
\be
T_{\mu\nu} = \frac{2}{\sqrt{-g}}\, \frac{\delta S}{ \delta g^{\mu\nu}}. 
\label{T-mu-nu}
\ee
Correspondingly for the complex field
\be 
\begin{split}
T^{(c)}_{\mu\nu} & = (\partial_\mu \phi_c^*) (\partial_{\nu}\,\phi_c) + (\partial_\nu \phi_c^*) (\partial_{\mu}\,\phi_c) - g_{\mu\nu} \left(g^{\alpha\beta}\partial_\alpha\phi_c^*\,{\partial_\beta \phi}_c - m^2 |\phi_c|^2\right) +  \\
& + \xi \left(2R_{\mu\nu} - g_{\mu\nu}\, R\right)\,|\phi_c|^2 - 2\xi\left(D_\mu D_\nu - g_{\mu\nu} D^2 \right)|\phi_c|^2,
\label{t-mu-nu-c}
\end{split}
\ee
where $D_\mu$ is the covariant derivative in metric \eqref{FLRW}. The trace of this tensor is
\be
T^{(c)\,\mu}_{\mu} = 2(6\xi-1) \partial_\mu \phi_c^* \partial^\mu \phi_c + 2\xi (6\xi-1) R |\phi_c|^2 + 4 (1-3 \xi) m^2 |\phi_c|^2. 
\label{trace t-c}
\ee
Note that for $\xi = 1/6$ and $m = 0$ the trace vanishes.

For the real field $\phi_r$ the energy--momentum tensor has the same form with twice smaller coefficients.

Fields $\phi$'s enter the equation of motion for $R$ \eqref{D2-R-1} via the trace of their energy--momentum tensors. Taking quantum average of $ T^\mu_\mu$ over background "filled" by classical scalaron field $R$, but devoid of $\phi$-particles, we can obtain equation for $R$ with an account for particle production. As we see in what follows, in the particular case of harmonic oscillations of the scalaron, particle production can be described by the simple term $\Gamma \dot R/ 2$.

\subsection{Decay into a pair of minimally coupled massless scalars \label{ss-dec-min-scal}}

The scalaron decay width into two massless (or very low mass) scalar bosons was calculated in~\cite{AAS-R2,Vilenkin_1985,EA-AD-LR}. Here we follow our paper~\cite{EA-AD-LR}, where another approach was used based on papers~\cite{Dolgov_Hansen,AD-Freese}, which allows to derive closed equation for an arbitrary time evolution of the source field (in our case the scalaron, $R(t)$), while the traditional methods are valid only for the harmonic oscillations of the source. 

According to Eq.~\eqref{A-phi-compl} the action for the complex massless scalar field with minimal coupling to gravity has the form:
\begin{equation}
S_c^{(00)} [\phi_c] = \int d^4x\,\sqrt{-g}\, g^{\mu\nu}\partial_\mu\phi_c^*\,\partial_\nu\phi_c 
\label{S-compl-00}
\end{equation}
and leads to the following equation of motion:
\begin{equation}
\ddot \phi_c + 3H\dot\phi_c - \frac{1}{a^2}\Delta\phi_c = 0.
\label{d2-phi-00}
\end{equation}

It is convenient to study particle production in terms of the conformally rescaled field, and the conformal time defined according to the equations:
\be
\chi_c = a(t)\phi_c, \quad d\eta = dt/a(t). 
\label{conf-trans-1}
\ee
The FLRW metric \eqref{FLRW} in conformal time is transformed into
\be
ds^2 = a^2 (\eta) \left(d\eta^2 - \delta_{ij} dx^i dx^j \right).
\label{FLRW-c}
\ee
The last factor in brackets is equal to the flat Minkowski metric. Such metrics are called conformally flat.

The curvature scalar in metrics \eqref{FLRW} and \eqref{FLRW-c} is expressed through the scale factor as
\be 
R = -6 \left( \dot H +2 H^2 \right) = -6 a''/a^3,
\label{R-def}
\ee
here and below prime denotes derivative with respect to conformal time.

The equation of motion for the conformally rescaled field $\chi_c$ takes the form:
\be
\chi_c''-\Delta\chi_c+\frac{1}{6}\,a^2R\,\chi_c = 0\,,
\label{chi-diprime-1}
\ee
while action \eqref{S-compl-00} turns into
\be
S_c^{(00)}[\chi_c] = \int d\eta\,d^3x\,\left(\chi_c'^*\chi_c' - \bm\nabla \chi_c^* \bm\nabla \chi_c - \frac{a^2R}{6}|\chi_c |^2\right),
\label{S-c-chi-00}
\ee
where $\bm\nabla $ is three-dimensional gradient in flat space.

Equation \eqref{D2-R-1}, which describes the scalaron evolution, can now be written as:
\be \nonumber
R'' + \frac{2a'}{a} R'&+&a^2 M_R^2 R = \\ 
&=& \frac{16\pi}{a^2} \frac{M_R^2}{M_\text{Pl}^2} \left[ \chi_c'^* \chi_c' - \bm\nabla \chi_c^* \bm\nabla \chi_c + \frac{a'^2}{a^2}|\chi_c|^2-\frac{a'}{a}(\chi_c^*\chi_c'+\chi_c'^*\chi_c)\right]\!.
\label{R-diprime}
\ee

Our aim is to derive a closed equation for $R$ taking the average value of the $\chi$-dependent quantum operators in the r.h.s. of Eq.~\eqref{R-diprime}, in presence of classical curvature field $R(\eta)$. The consideration essentially repeat those of~\cite{Dolgov_Hansen}, where the equation was derived in one-loop approximation.

Equation \eqref{chi-diprime-1} can be transformed into the following integro-differential equation convenient for perturbative solution:
\be
\label{formal_sol}
 \chi_c (x) = \chi_c^{(0)}(x)-\frac{1}{6}\int d^4y\,G(x,y)\,a^2(y)R(y)\chi_c(y)\equiv \chi_c^{(0)}(x)+\delta \chi_c (x)\,,
\ee
where $\chi_c^{(0)}$ satisfies the free equation $ \chi''_c - \Delta \chi_c = 0$ and the massless Green's function is
\begin{equation}
\label{green_func}
G(x,y) = G(x-y) = \frac{1}{4\pi|\mathbf x-\mathbf y|} \delta\big((x_0 - y_0) - |\mathbf x -\mathbf y| \big) \equiv \frac{1}{4\pi \Delta r}\delta(\Delta\eta - \Delta r).
\end{equation}
Here $\Delta \eta = x_0-y_0$ and $\Delta r = |\mathbf x-\mathbf y|$. Since $\Delta r = |\mathbf x-\mathbf y| \geq 0$, the condition $\Delta \eta \geq 0$ is also to be fulfilled.

The free field $\chi_c^{(0)}$ is quantized in the usual way:
\begin{equation}
\label{quantiz_chi}
\chi_c^{(0)}(x) = \int\frac{d^3k}{2 E_{\mathbf k} (2\pi)^3} \left[\hat a_{\mathbf k}\,e^{-ikx} + \hat b^\dagger_{\mathbf k}\,e^{ikx}\right],
\end{equation}
where $x^\mu = (\eta,\mathbf x)$, $k^\mu = (E_{\mathbf k}, \mathbf k)$, $E_{\mathbf k}^2 - {\mathbf k}^2 = 0$, and $\hat a_{\mathbf k}$ is the annihilation operator for particles, while $\hat b^\dagger_{\mathbf k}$ is the creation operator for antiparticles. The creation--annihilation operators satisfy the commutation relations:
\be
\label{commutator}
\left[\hat a_{\mathbf k}, \hat a_{\mathbf k'}^\dagger \right] = 2E_{\mathbf k} (2\pi)^3\,\delta^{(3)}(\mathbf k - \mathbf k'),
\ee
and analogously for $\hat b_{\mathbf k}$. All other commutators vanish.

The particle production effects are assumed to weakly perturb the free solution, so Eq.~\eqref{formal_sol} can be solved in the first order perturbation approximation as
\begin{equation}
\label{approx_chi}
 \chi_c(x)\simeq \chi_c^{(0)}(x)-
 \frac{1}{6}\int d^4y\,G(x-y)\,a^2(y)R(y)\chi_c^{(0)}(y)\equiv\chi_c^{(0)}(x)+\chi_c^{(1)}(x).
\end{equation}
Now we calculate the vacuum expectation values of the various terms in the r.h.s. of Eq.~\eqref{R-diprime}, keeping only the contribution from the terms linear in $\chi_c^{(1)}$. The terms of zero order which are bilinear in $\chi^{(0)}$ and its derivatives have nothing to do with particle production and can only change the parameters of the theory through the renormalization procedure. 

We need to calculate the products of the quantum operators of the kind:
\be
\langle \chi_c^{(0)} (x) \chi_c^{(1)*} (x) \rangle = -\frac{1}{6} \int d^3 y\, dy_0\, G(x-y) a^2 (y_0) R(y_0) \langle \chi_c^{(0)} (x) \chi_c^{(0)*} (y) \rangle, 
\label{chi-chi-average}
\ee
where $dy_0 \equiv d\eta_y$ is the time component corresponding to the space coordinate $dy$. 

The vacuum expectation values of the creation/annihilation operators are
\be
\langle \hat b_{{\mathbf k}_1}^\dagger \hat b_{{\mathbf k}_2} \rangle = 0, \qquad
\langle \hat a_{{\mathbf k}_1} \hat a^\dagger_{{\mathbf k}_2} \rangle = 2E_{\mathbf k} (2\pi)^3\,\delta^{(3)}(\mathbf k_1 - \mathbf k_2),
\label{vac-avar}
\ee
where in the last equation we have used commutator \eqref{commutator}. 

Now using expansion \eqref{quantiz_chi} we find
\be
\langle \chi_c^{(0)} (x) \chi_c^{(1)*} (x) \rangle = -\frac{1}{6} 
\int \frac{d^3 k }{2 E_{\mathbf k} (2\pi)^3} d^3y\, dy_0\, G(x-y) a^2(y_0) R(y_0)\,
e^{- i E_{\mathbf k} (x_0 - y_0)+i {\mathbf k} ({\mathbf x} - {\mathbf y})}. \qquad
\label{aver-1}
\ee
Let us first integrate over angles in $d^3 k = E_{\mathbf k}^2 dE_{\mathbf k} d(\cos\theta) d\phi$: 
\be
\langle \chi_c^{(0)} (x) \chi_c^{(1)*} (x) \rangle = -\frac{1}{48\pi^2} \int d^3y\, dy_0\, dk \frac{e^{ik\Delta r} - e^{-ik\Delta r}}{i \Delta r} e^{-ik\Delta \eta}\,G(x-y) a^2(y_0) R(y_0). \quad
\label{int-angle}
\ee
For brevity we used the notation $k = E_k = |\mathbf k|$. 

Next we integrate over $d^3 y = d^3 \Delta r$ using the delta-function in the expression for the Green function \eqref{green_func} and obtain:
\be
\langle \chi_c^{(0)} (x) \chi_c^{(1)*} (x) \rangle = \frac{i}{48\pi^2} \int dy_0\, dk\, \left(1 - e^{-2ik\Delta \eta} \right) a^2(y_0) R(y_0).
\label{int-dr}
\ee
Therefore, 
\be
\langle |\chi_c(x)|^2 \rangle \simeq \langle 2\,\mathrm{Re} (\chi_c^{(0)}\chi_c^{(1)*}(x))\rangle = 
-\frac{1}{24\pi^2}\int dy_0\, dk\,\mathrm{Re}\left(ie^{-2ik\Delta \eta} \right) a^2(y_0) R(y_0).
\ee

The integral over $dk$ can be taken according to the equation:
\be 
\int_0^\infty dk\,e^{i\alpha k} = \pi \delta(\alpha) + i {\cal P} \left(\frac{1}{\alpha} \right),
\label{int-exp-1}
\ee
so we arrive finally at 
\be
\langle |\chi_c(x)|^2 \rangle \simeq -\frac{1}{48\pi^2}\int dy_0\, a^2(y_0) R(y_0){\cal P}\left(\frac{1}{\Delta \eta}\right).
\ee
The upper limit of integration is imposed by the condition $\Delta \eta \geq 0$, see Eq.~\eqref{green_func}.

The dominant contribution to the particle production comes from the first term in the r.h.s. 
of Eq.~\eqref{R-diprime}. So we have to calculate the expectation value:
\be
\langle (\chi^{(0)}_c (x))' (\chi^{(1)}_c (x))'^* \rangle = 
-\frac{1}{6} \int d^3 y\, dy_0\, \frac{\partial G(x-y)}{\partial x_0}\, a^2 (y_0) R(y_0) \langle (\chi^{(0)}_c (x))' (\chi^{(0)}_c (y))'^* \rangle. \qquad
\label{chi-chi-average}
\ee
Taking into account that $\partial G(x-y)/\partial x_0 = -\partial G(x-y)/\partial y_0$ and
integrating by part over $dy_0$ we get
\be
\begin{split}
\langle (\chi^{(0)}_c (x))' (\chi^{(1)}_c (x))'^* \rangle & = -\frac{1}{6} \int d^3 y\, dy_0\, G(x-y) a^2(y_0)\times  \\
& \times \left[R'(y_0) \langle (\chi^{(0)}_c (x))' \chi^{(0)*}_c (y) \rangle + 
R(y_0)\langle (\chi^{(0)}_c (x))' (\chi^{(0)}_c (y))'^* \rangle 
\right],
\end{split}
\label{d-y0}
\ee
where the derivative of $a^2 (y_0) $ is neglected because it is slowly varying function of time.

In complete analogy with the calculations made above we find
\be
\begin{split}
\langle (\chi^{(0)}_c (x))' (\chi^{(1)}_c (x))'^* \rangle & = 
-\frac{1}{6} \int \frac{d^3 k }{2 E_{\mathbf k} (2\pi)^3}\, d^3y\, dy_0\, \,G(x-y) a^2(y_0) \times  \\
& \times e^{- i E_{\mathbf k} (x_0 - y_0)+i {\mathbf k} ({\mathbf x} - {\mathbf y})} \left(-i E_{\mathbf k} R' + E_{\mathbf k}^2 R \right).
\end{split}
\label{aver-2}
\ee
Integration over directions of $d^3 k$ leads as above to
\be
\begin{split}
\langle (\chi^{(0)}_c (x))' (\chi^{(1)}_c (x))'^* \rangle & = 
-\frac{1}{48\pi^2} \int d^3y\, dy_0\, dk\, \frac{e^{ik\Delta r} - e^{-ik\Delta r}}{i \Delta r}\, e^{-ik\Delta \eta} \times  \\
& \times G(x-y) a^2 \left(-i E_{\mathbf k} R' + E_{\mathbf k}^2 R \right).
\end{split}
\ee
After integration over $d^3y$ with $G(x-y)$ given by Eq.~\eqref{green_func} we arrive at
\be
\langle (\chi^{(0)}_c (x))' (\chi^{(1)}_c (x))'^* \rangle = 
\frac{i}{48\pi^2} \int dy_0\, dk \left(1 - e^{-2ik\Delta \eta} \right) a^2(y_0) \left(-i k R' + k^2 R\right).
\label{int-dr-2}
\ee
Using equations 
\[
2ik \exp(-2ik\Delta\eta) = \partial_{y_0} \exp (-2ik\Delta\eta), \quad 
4k^2 \exp (-2 ik \Delta \eta) = -\partial^2_{y_0} \exp (-2 ik \Delta \eta)
\]
and integrating by parts we obtain
\be
{\langle |\chi_c'(x)|^2 \rangle \simeq -\frac{1}{192\pi^2}\int dy_0\, a^2(y_0) R''(y_0){\cal P}\left(\frac{1}{\Delta \eta}\right)}.
\label{int-by-parts-2}
\ee

In a similar way we get
\be
\langle |\bm\nabla \chi_c(x)|^2 \rangle \simeq \frac{1}{192\pi^2}\int dy_0\, a^2(y_0) R''(y_0){\cal P}\left(\frac{1}{\Delta \eta}\right),
\ee
\be
\langle (\chi_c^*(x)\chi_c'(x)+\chi_c'^*(x)\chi_c(x))\rangle \simeq
-\frac{1}{48\pi^2}\int dy_0\, a^2(y_0) R'(y_0){\cal P}\left(\frac{1}{\Delta \eta}\right).
\ee

Inserting these expressions into \eqref{R-diprime}, we obtain a closed integro-differential equation for $R$, which we will transform into ordinary differential equation for harmonic oscillations of $R$ neglecting the slow power law decrease of its amplitude at the scale of very fast oscillations. 

By the same reason the scale factor, $a(t)$, varies very little during many oscillation times, $\omega^{-1} = M_R^{-1}$. Thus, we expect that $d\eta/\eta\sim dt/t$ and that the dominant part in the integrals in \eqref{approx_chi} is given by derivatives of $R$, since $R'\sim \omega R+t^{-1}R\simeq\omega R$, because $\omega t\gg 1$. So the dominant contribution of particle production is given by expression
\be
\langle (|\chi_c'(x)|^2-|\bm\nabla \chi_c(x)|^2) \rangle \simeq -\frac{1}{96\pi^2}\int^\eta_{\eta_0}d\eta_1\,\frac{a^2(\eta_1)R''(\eta_1)}{\eta-\eta_1}\,,
\label{chi_prime_squared}
\ee
and is reduced to
\begin{equation}
\label{R_with_back_reaction_approx}
\ddot R+3H\dot R+M^2_R R \simeq -\frac{1}{6\pi} \frac{M_R^2}{M_\text{Pl}^2}\frac{1}{a^4} \int_{\eta_0}^\eta d\eta_1\,\frac{a^2(\eta_1)R''(\eta_1)}{\eta-\eta_1} \simeq 
-\frac{1}{6\pi} \frac{M_R^2}{M_\text{Pl}^2}\int_{t_0}^t dt_1\,\frac{\ddot R(t_1)}{t-t_1}\,.
\end{equation}
The equation is naturally non-local in time since the effect of particle production depends upon all the history of the system evolution. 

Rigorous determination of the decay width of the scalaron is described in~\cite{EA-AD-LR}. Here we present it in a simpler and intuitively clear way. We will look for the solution of Eq.~\eqref{R_with_back_reaction_approx} in the form:
\be 
R = R_\text{amp} \cos (\omega t + \theta) \exp( -\Gamma t/2),
\label{R-of-gamma}
\ee
where $R_\text{amp}$ is the slowly varying amplitude of $R$-oscillations, $\theta$ is a constant phase depending upon initial conditions, and $\omega$ and $\Gamma$ are to be determined from Eq.~\eqref{R_with_back_reaction_approx}. The term $3 H \dot R$ is not essential in the calculations presented below and will be neglected. The exponent is taken equal to $\Gamma t/2$ so the scalaron energy density would decrease as $\exp (-\Gamma t)$.

Assuming that $\Gamma$ is small, so the terms of order of $\Gamma^2$ are neglected and treating the r.h.s. of Eq.~\eqref{R_with_back_reaction_approx} as perturbation we obtain:
\be
\begin{split}
& \left[\left(- \omega^2 + M_R^2 \right) \cos (\omega t + \theta) + \Gamma \omega \sin(\omega t + \theta) \right] e^{-\Gamma t/2} =  \\
& = \frac{1}{6\pi} \frac{\omega^2 M_R^2}{M_\text{Pl}^2}\, e^{-\Gamma t/2} \int_0^{t-t_0} \frac{d\tau}{\tau} 
\left[\cos(\omega t +\theta) \cos (\omega \tau) + \sin(\omega t +\theta) \sin (\omega \tau) \right].
\label{def-Gamma}
\end{split}
\ee
The first, logarithmically divergent, term in the integrand leads to mass renormalization and can be included into physical $M_R$, while the second term is finite and can be analytically calculated at large upper integration limit $\omega t$ according to the well-known integral
\be
\int_0^\infty \frac{d\tau}{\tau} \sin (\omega \tau) = \frac{\pi }{2}.
\label{int-exp}
\ee

Comparing the l.h.s. and r.h.s. of Eq. \eqref{def-Gamma} we can conclude that $\omega = M_R$ and the width of the scalaron decay into a pair of ``charged'' (complex) {minimally coupled massless} scalars is
\be
\Gamma_c {(\xi = 0, m = 0)} = \frac{M_R^3}{12 M_\text{Pl}^2}.
\label{Gamma-c}
\ee

To find the decay width of the scalaron into a pair of neutral (real) scalars we can apply the following arguments. The scalaron interacts with the trace of the energy--momentum tensor of the complex scalar, $\phi_c$, {see \eqref{trace t-c}, and the} neutral one, 
$\phi_0$, so the interaction term 
{is proportional to}
\be
\left( 2 \partial_\mu \phi_c\, \partial^\mu \phi^*_c + \partial_\mu \phi_0\, \partial^\mu \phi_0 \right) = \left( \partial_\mu \phi_1\, \partial^\mu \phi_1 + \partial_\mu \phi_2 \, \partial^\mu \phi_2 + \partial_\mu \phi_0\, \partial^\mu \phi_0 \right) 
{ \,=\, \partial_\mu \bm\phi\, \partial^\mu \bm\phi } ,
\label{L-00}
\ee
where{ $\phi_c = (\phi_1 +i \phi_2)/\sqrt{2}$, $\phi_c^* = (\phi_1 -i \phi_2)/\sqrt{2}$, and $\bm\phi = (\phi_1, \phi_2, \phi_0)$ is the isotopic vector.}
Isotopic invariance leads to equal number of the produced $\phi^+$, $\phi^-$, and $\phi_0$ particles, as is well known to be realized in pion physics. Hence the width of the scalaron decay into a pair of neutral (real) identical particles should be evidently twice smaller, than the width of the decay into the charged ones and so
\be
\Gamma_r {(\xi = 0, m = 0)} = \frac{M_R^3}{24 M_\text{Pl}^2}.
\label{Gamma-r}
\ee
The latter result agrees with those presented e.g. in~\cite{Vilenkin_1985, gorb-pan}. However, is it twice larger than the width of the scalaron decay into two real massless scalars calculated in paper~\cite{faulkner}, Eq. (76).

\subsection{Decay into a pair of minimally coupled massless scalars, another method \label{s-scal-decay-2}} 

Now we calculate the rate of the scalaron decay into the same channel, as is studied in the previous subsection, in a different way dealing with the energy loss of the scalaron into the produced particles. To this end we will use the equation of motion of the decay products \eqref{chi-diprime-1} and calculate the energy density of particles $\chi$ and anti-$\chi$ created by the oscillating gravitational field of the scalaron per unit time, $\dot \rho_\chi$. Then we compare it to the energy density of the canonically normalized scalaron field~\eqref{Phi}.

In what follows we use for $R$ the solution \eqref{R-of-t}: $R = -4 M_R \cos (M_R t + \theta)/t$. For this $R$ the energy density of $\Phi$, as follows from Eq. \eqref{rho-R-1}, is equal to
\be
\rho_\Phi = \frac{\dot\Phi^2 +M_R^2 \Phi^2}{2} = 
\frac{M_\text{Pl}^2 (\dot R^2 + M_R^2 R^2)}{96 \pi M_R^4} \approx \frac{M_\text{Pl}^2}{6 \pi t^2} \quad {(\text{for}\ M_R t\gg 1)}.
\label{rho-R}
\ee
Note that this is formally equal to the critical energy density for matter dominated universe.

Due to the energy conservation $\dot \rho_\chi + \dot \rho_{\bar\chi} = 2\dot \rho_\chi = - \dot \rho_\Phi $.
So for the rate of the energy dissipation of the scalaron we find:
\be
\Gamma = -\frac{\dot \rho_\Phi}{\rho_\Phi } = \frac{2 \dot \rho_\chi }{ \rho_\Phi},
\label{Gamma-R-3}
\ee
where $ \dot \rho_\chi $ is calculated along the standard lines of particle production theory in external time-dependent fields, see e.g.~\cite{Grib-pp,AD-DK,CB-AD}.

According to the Parker theorem~\cite{parker1, parker2} massless particles are not created by conformally flat FLRW metric. This is fulfilled for massless spin-1/2 fermions, massless gauge boson (up to conformal anomaly~\cite{AD-conf-anom}), but is not always true for scalar bosons, because the latter are conformally invariant only for $\xi = 1/6$.

The particle production in conformally flat FLRW background is convenient to study in terms of conformal time as described above in Eqs.~\eqref{conf-trans-1}--\eqref{S-c-chi-00}. In what follows we closely follow book~\cite{CB-AD}. The quantum field operator describing the created particles is assumed to satisfy the equation:
\be
\chi'' - \Delta \chi + f(\eta) \chi = 0.
\label{chi-di-prime-2}
\ee
We omit here subindex $c$ at $\chi_c$ {for brevity}, because only complex field $\chi$ is considered in this section. Field $\bar \chi$ satisfies the Hermitian conjugate equation. The function $f(\eta)$ is a classical external field producing quanta of $\chi$ and anti-$\chi$ particles. 

The amplitude of production of a pair of $\chi$ and anti-$\chi$ bosons with momenta $\mathbf k_1$ and $\mathbf k_2$, respectively, is equal to the matrix element of the interaction term between vacuum and the corresponding particle--antiparticle state:
\be
A(k_1, k_2) = \int d\eta\, d^3 x\, f(\eta) \langle \mathbf k_1, \mathbf k_2 | \chi^\dagger (\eta, \textbf{x}) \chi (\eta, \textbf{x}) |{0}\rangle.
\label{A-k1-k2}
\ee
Recall that we {use}
conformal time $d\eta = dt/a(t)$ and conformally transformed field $\chi = a \phi$.

The quantum field operators are expanded in terms of the creation/annihilation ope\-rators as given by Eq.~\eqref{quantiz_chi}. The ``bra'' state of the produced pair of $\chi$ and $\bar \chi$ quanta is defined in terms of these operators as
\be
\langle \mathbf k_1, \mathbf k_2| = \langle {0}| \hat a_{\mathbf k_1} \hat b_{\mathbf k_2}. 
\label{produced-pair}
\ee
Now keeping in mind that the annihilation operator acting on the vacuum state to the right annihilates the state, $\hat a_{\mathbf k} | 0 \rangle = 0$, and correspondingly $\langle 0 | \hat{a}^\dagger_{\mathbf k} = 0$, so applying commutation relation \eqref{commutator} we obtain:
\be
\begin{split}
A (k_1, k_2) & = \int d\eta\, d^3 x\, d\tilde k\, d\tilde k' \langle 0 | \hat a_{\mathbf k_1} \hat b_{\mathbf k_2} \hat a^\dagger_{\mathbf k} \hat b^\dagger_{\mathbf k'} | 0 \rangle f(\eta) e^{i\eta (E_\textbf{k} + E_\textbf{k'}) - i \textbf{x} (\textbf{k} + \textbf{k'})} =  \\
& = (2\pi)^3 \delta^{(3)} (\mathbf k_1 + \mathbf k_2 ) \int d\eta\, f(\eta) e^{i\eta(E_{\mathbf k_1}+E_{\mathbf k_2})},
\end{split}
\ee
where $d\tilde k = d^3 k /(2E_\textbf{k} (2\pi)^3 )$ and $E_\textbf{k} = |\textbf{k}|$. Hence
\be
\left|A (k_1,k_2)\right|^2 = (2\pi)^3 V \delta^{(3)} (\mathbf k_1 + \mathbf k_2 ) \left| \int d\eta\, f(\eta) e^{i\eta(E_{\mathbf k_1}+E_{\mathbf k_2})}\right|^2,
\label{A2}
\ee
here the following identities are used: 
\be
\int d^3 x\, e^{-i \textbf{k x}} = (2\pi)^3 \delta^{(3)} ( \textbf{k}) \quad {\rm and} \quad \delta^{(3)} (\textbf{k = 0}) = V/(2\pi)^3, 
\label{int-delta}
\ee 
where $V$ is the total space volume.

The Fourier transform of the source can be simply calculated for the case of harmonic oscillations
\be
f(\eta) = f_0 \cos (\omega \eta) = \frac{f_0}{2} \left(e^{i\omega \eta} + e^{-i\omega \eta} \right),
\label{f-harm}
\ee
where $f_0$ is slowly changing compared to $\cos (\omega \eta)$ function of $\eta$, so we consider it as a constant. Taking into account that the energy of the created particles should be positive, $E_{\mathbf k_{1,2}}>0$, we obtain
\be 
\left| \int d\eta\, f(\eta) e^{i \eta (E_{\mathbf k_1} + E_{\mathbf k_2})} \right|^2 = 
\frac{\pi}{2} f_0^2 \delta (\omega - E_{\mathbf k_1} - E_{\mathbf k_2}) \Delta \eta,
\label{fourrer2}
\ee
where we used that $\delta (0) = \Delta \eta/(2\pi)$ and $\Delta \eta$ is the time duration of the process, presumably $\omega \Delta \eta \gg 1 $.

So the $\chi$-particle production rate per unit volume and unit conformal time is
\be
\begin{split}
\frac{d n_\chi^{(\text{tot})}}{d\eta} & = \frac{2 d n_\chi}{d\eta} = \frac{2}{V \Delta \eta} \int d^3 \tilde k_1 d^3 \tilde k_2 | A(k_1,k_2)|^2 =  \\
& = \pi (2\pi)^3 f_0^2 \int d^3 \tilde k_1\, d^3 \tilde k_2\, \delta^{(3)} (\mathbf k_1 + \mathbf k_2 ) \delta (\omega - E_{\mathbf k_1} - E_{\mathbf k_2} ) = \frac{f_0^2}{16 \pi}.
\end{split}
\label{dot-n-chi}
\ee

Now we have to express $f(\eta) = f_0 \cos (\omega \eta)$ through $R(t)$. Comparing Eq.~\eqref{chi-diprime-1} with \eqref{chi-di-prime-2} and using Eq.~\eqref{R-of-t} we conclude that
\be
f_0 \cos (\omega \eta) = - \frac{2 a^2 M_R}{3 t}\, \cos (M_R t + \theta).
\label{f-of-R}
\ee
Let us show that we can take $f_0 = 2 a^2 M_R/(3 t)$. Since Eq.~\eqref{f-of-R} holds only for $M_R t \gg 1$, oscillation period $\Delta t = 2 \pi/M_R $ is much less than absolute value of time, $\Delta t \ll t $. Then since the product $p_\mu dx^\mu$ is a scalar with respect to the general coordinate transformation and in our case $p_\mu$ has only one nonzero component, namely, $p_\mu = (M_R, \bm 0)$, we can conclude that $\omega d\eta = M_R dt $ and $\omega \Delta\eta \approx M_R \Delta t $ during oscillation period. Finally, choosing the initial value of conformal time $\eta = 0$ corresponding to the time $t$ at which $\cos (M_R t + \theta) = -1$ we obtain that
\be
f_0 = \frac{2 a^2M_R}{3t} . 
\label{f-of-R-old}
\ee
The r.h.s. of Eq. \eqref{dot-n-chi} is proportional to $a^4$. The same is true for its l.h.s. because $d/d\eta = a d/dt$ and $n_\chi \sim \chi \chi' \sim a^3 {n_\phi}$. So returning to $dn_\phi /dt $ we find that it does not depend upon $a$. Since the energy of a $\phi$-particle in the physical frame is equal to $M_R/2$, the time derivative of the energy density $\dot \rho_\phi$ lost to creation of $\phi$-particle is obtained from $\dot n_\phi$ by multiplication by $M_R/2$ and we find
\be
\dot \rho_\phi = \frac{M_R}{2}\, \dot n_\phi = \frac{M_R}{2} \frac{f_0^2}{32 \pi a^4} = \frac{M_R^3}{144 \pi t^2}.
\label{dot-rho-chi}
\ee

The energy density of the scalaron field \eqref{rho-R} is $M_\text{Pl}^2 / (6\pi t^2)$ and hence the width of the scalaron decay into a pair of ``charged" (complex) minimally coupled massless $\phi$-particles is
\be
\Gamma_c (\xi = 0, m = 0) = \frac{\dot \rho^{(\text{tot})}_\phi}{\rho_R} = \frac{2\dot\rho_\phi}{\rho_\Phi} = \frac{M_R^3}{12 M_\text{Pl}^2}.
\label{Gamma-2}
\ee
It agrees with result \eqref{Gamma-c} of the previous section.

As for the width of scalaron decay into a pair of neutral (real) minimally coupled massless $\phi$-particles, it is twice smaller than $\Gamma_c$:
\be
\Gamma_r (\xi = 0, m = 0) = \frac{ \Gamma_c }{2} = \frac{M_R^3}{24 M_\text{Pl}^2}.
\label{Gamma-r-2}
\ee


\subsection{Decay into conformally coupled massive scalars \label{s-scal-decay-2}} 

Let us consider now the case of conformally coupled decay products, i.e. Eq.~\eqref{ddot-phi} with $\xi = 1/6$ and $m \neq 0$, but still $m \ll M_R$, so the phase space suppression is not essential. 

In terms of the conformally rescaled field $\chi = a \phi$ and the conformal time $\eta$ Eq.~\eqref{ddot-phi} transforms into
\be
\chi'' - \Delta\chi + \left(\frac{1}{6} -\xi \right) a^2 R \chi + m^2 a^2 \chi = 0.
\label{chi-diprime-2}
\ee
Here prime means differentiation over $\eta$ and $ R = -6a''/a^3$. The temporal evolution of $R(t)$ is given by Eq.~(\ref{R-of-t}).

Therefore, the interaction leading to the particle production in the case of $\xi = 1/6$ has the form:
\be 
V = m^2 a^2.
\label{V?}
\ee
Using the solution~{\eqref{H-of-t}, }
\be 
H = \frac{\dot a}{a} = \frac{2}{3t} [1 + \sin (M_R t + \theta)],
\label{H-of-t-sol}
\ee
we find {for $M_R t \gg 1$ that}
\be
{a(t) = a_0\, t^{2/3} \exp \left\{ - \frac{2 \cos (M_R t + \theta)}{3 M_R t}\right\} \approx a_0\, t^{2/3}, }
\label{a-of-t-sol}
\ee
and consequently,
\be
V(t) = m^2 a^2(t) \approx m^2 a_0^2\, t^{4/3} \left\{1 - \frac{4\cos (M_R t + \theta)}{3 M_R t} \right\}\ \to\ -\frac{4 m^2 a^2 }{3 M_R t} \cos (M_R t + \theta), \quad
\label{a-of-t-expad}
\ee
since the term $m^2 a_0^2\, t^{4/3}$ has nothing to do with particle production and can be omitted. 
 
Comparing it with the expression for $R(t)$ \eqref{R-of-t} and Eq.~\eqref{dot-n-chi} we can conclude that the energy release from $\phi$ decay into the primeval plasma, if the decay is induced by {$m^2 a^2(t)$ term}, is equal to
\be
\dot \rho_\phi = \frac{M_R}{2}\, \dot n_\phi = \frac{M_R}{2} \frac{V_0^2}{32 \pi a^4} = 
\frac{M_R}{64 \pi a^4} \left(\frac{4 m^2 a^2}{3 M_R t} \right)^2 = \frac{m^4}{36 \pi M_R t^2},
\label{dot-rho-0-0}
\ee
and the {width of scalaron decay into a pair of ``charged" (complex) conformally coupled massive scalars is}
\be
\Gamma_c (\xi = 1/6, m \neq 0) = \frac{\dot \rho^{(\text{tot})}_\phi}{\rho_R} = \frac{2 \dot \rho_\phi}{\rho_\Phi} = 
{ \frac{m^4}{18 \pi M_R t^2} \frac{6 \pi t^2}{M_\text{Pl}^2} =\, }
\frac{m^4}{3 M_R M_\text{Pl}^2}.
\label{Gamma-xi-m}
\ee
The width of scalaron decay into a pair of neutral (real) conformally coupled massive scalars is twice smaller:
\be
\Gamma_r (\xi = 1/6, m \neq 0) = \frac{m^4}{6 M_R M_\text{Pl}^2}.
\label{Gamma-r-m}
\ee
This result coincides with that of~\cite{Vilenkin_1985}.

\section{Decays into fermions \label{s-fermions}}

Let us first consider the production of fermions in the Minkowski space--time by scalar field, $\phi$, which interacts with fermions, $\psi$, according to the Lagrangian
\be
L_{\phi \psi \psi} = g \phi \bar \psi \psi,
\label{L-phi-psi}
\ee
where $g$ is a dimensionless coupling constant. The field $\phi$ is supposed to be harmonically oscillating:
\be
\phi (t) = \phi_0 \cos (\Omega t).
\label{phi-osc}
\ee
The production amplitude is equal to the matrix element {of $L_{\phi \psi \psi}$} between vacuum and fermion--antifermion state with momenta $p_1$ and $p_2$:
\be
A(p_1, p_2) = g \int d^4 x\, \phi (t) \langle p_1 ,p_2 |\bar\psi (x) \psi(x) | 0 \rangle .
\label{A-p1-p2}
\ee
The fermion operator wave functions are decomposed in terms of creation--annihilation operators as
\be
\psi (x) = \int \frac{d^3p}{(2 \pi )^3 (2 E_\textbf{p})} \, 
\sum_s \left(\hat{a}^s_\textbf{p} u^s(p) e^{-ipx} + \hat{b}^{s \dag}_\textbf{p} v^s(p) e^{ipx}\right),
\label{qmferm}
\ee
where $\hat{a}^s_\textbf{p}$ and $\hat{a}^{s \dag}_\textbf{p}$ ($\hat{b}^s_\textbf{p}$ and $\hat{b}^{s \dag}_\textbf{p}$) are annihilation and creation operators of (anti)fermions with momentum $\textbf{p}$ and spin $s$, which obey the {anticommutation} relations:
\be
{ \{ \hat{a}^r_\textbf{p}, \hat{a}^{s \dag}_\textbf{q} \} = \{ \hat{b}^r_\textbf{p}, \hat{b}^{s \dag}_\textbf{q} \} = 
2E_\textbf{p} (2 \pi)^3 \delta^{(3)} (\textbf{p} - \textbf{q}) \delta^{r s}, }
\label{anti-com}
\ee
where {$E_\textbf{p} = \sqrt{\textbf{p}^2 + m^2 } \approx |\textbf{p}|$} if the mass is small in comparison with {$|\textbf{p}|$}.
The summation over spins 
is done with the usual relations:
\[ 
{ \sum u^s(p) \bar{u}^s(p) = \slash \!\!\!p + m \quad \text{and} \quad \sum v^s(p) \bar{v}^s(p) = \slash \!\!\!p - m. }
\]
The vacuum state is defined as zero-particle state, i.e. such that the annihilaiton operator kills it, 
{$\hat{a}^s_\textbf{p} |0\rangle = 0$.} The vacuum state is normalized as {$\langle 0|0 \rangle = 1$.} The two-fermion state in Eq.~\eqref{A-p1-p2} is defined as
\be
{ \langle p_1, p_2 |= \langle 0| \hat{a}^s_{\textbf{p}_1} \hat{b}^r_{\textbf{p}_2}. }
\label{two-ferm} 
\ee
Using the operator expansions \eqref{qmferm} and making proper commutations according to \eqref{anti-com} to reduce 
the amplitude to vacuum-to-vacuum transition we arrive at
\be
{ A(p_1, p_2) = (2\pi)^3 \delta^{(3)} ( \textbf{p}_1 + \textbf{p}_2) g \tilde\phi (E_1+ E_2) \bar u(p_1) v (p_2), }
\label{A-int1} 
\ee
where 
\be
{ \tilde\phi (\omega) = \int dt\, \phi (t) e^{i \omega t} }
\label{phi-tilde}
\ee
is the {Fourier transform of $\phi(t)$}. 

Next we calculate the amplitude squared taking into account that the trace 
$|\bar u(p_1) v (p_2) |^2$, which appears after summation {over the spin states} of the produced fermions is 
\be
{\mathrm{Tr} \left[ (\slash \!\!\!p_1 + m_\psi) (\slash \!\!\!p_2 - m_\psi)\right] = 4 \left[ (p_1 p_2) - m_\psi^2 \right]}
\label{trace}
\ee
and
\be
{ \left[(2\pi)^3 \delta^{(3)} ( \textbf{p}_1 + \textbf{p}_2) \right]^2 = (2\pi)^3 \delta^{(3)} ( \textbf{p}_1 + \textbf{p}_2) V, }
\label{delta-square}
\ee
where $V$ is the total space volume. Hence we obtain the following expression: the amplitude of the creation of fermion--antifermion pair summed over the spin states of the created fermions:
\be
{ | A(p_1, p_2) |^2 = 4\, (2\pi)^3 \delta^{(3)} ( \textbf{p}_1 + \textbf{p}_2) V | g \tilde \phi (E_1 + E _2)|^2 \left[ (p_1 p_2) - m_\psi^2 \right]. }
\label{A-2}
\ee

The number density of the produced fermions equals to
\be
n_\psi = \frac{1}{V} \int \frac{d^3 p_1}{2 E_1 (2 \pi)^3} \frac{d^3 p_2}{2 E_2 (2 \pi)^3} |A(p_1, p_2)|^2 = \frac{|g|^2 }{\pi^2} 
\int dE\, E^2 \big| \tilde \phi (2 E) \big|^2 \left(1-\frac{m_\psi^2}{E^2}\right)^{3/2}. \qquad
\label{n}
\ee 
Making the Fourier transformation of $\phi (t)$, Eq.~\eqref{phi-osc}, we find
\be 
\tilde \phi (2 E) = \pi \phi_0 \left[\delta (2 E + \Omega) + \delta (2 E - \Omega) \right] = \pi \phi_0\, \delta (2 E - \Omega),
\label{tilde-phi-1}
\ee
since $E > 0$ and $\Omega > 0$.

The square of this function is equal to
\be
{ \big| \tilde \phi (2 E) \big|^2 = \pi^2 |\phi_0|^2 \delta (2 E - \Omega) \delta (0) = \frac{\pi}{2} |\phi_0|^2 \delta (2 E - \Omega) \Delta t, }
\label{tilde-]phi-2}
\ee
where {$\delta (0) = \Delta t/(2 \pi)$ and} $\Delta t$ is the process duration. Hence
\be
\dot n_\psi = \frac{n_\psi}{\Delta t} = \frac{|g \phi_0|^2}{2 \pi} \int dE\, E^2 {\left(1-\frac{m_\psi^2}{E^2}\right)^{3/2} 
\delta (2E - \Omega)} = \frac{|g \phi_0|^2 \Omega^2}{16 \pi} \left(1-\frac{4 m_\psi^2}{\Omega^2}\right)^{3/2}. \quad
\label{dot-n-ferm}
\ee
 
To calculate fermion production by oscillating {curvature} we need to go to conformal time  and conformal variables, and to make the substitution (see {Appendix}):
\be 
g \phi = g \phi_0 \cos (\Omega t)\ \to\ m_\psi a_\text{osc} (\eta) = 
\frac{2 m_\psi a_\text{bg}}{3 M_R t } \cos \left(M_R a_\text{bg} \eta \right).
\label{sub-to-eta} 
\ee
Using Eq.~\eqref{dot-n-ferm} we find the fermion density production per unit conformal time as:
\be
\tilde n'_\psi \approx \frac{1}{16 \pi} \left( \frac{2 m_\psi a_\text{bg}}{3 M_R t } M_R a_\text{bg} \right)^2 \approx 
\frac{m_\psi^2 a^4}{36 \pi t^2},
\label{tilde-n-prime}
\ee
{where we take into account that $\Omega = M_R a_\text{bg} \gg m_\psi$}, and in the denominator the usual time $t$ is kept, because it cancels out in the final expression for $\Gamma$.
 
Now we go to the physical time $dt = a\, d\eta$ and to physical number density $n_\psi = a^3 \tilde n_\psi$, as it is formally clear from the definition of conformally transformed spinors, Eq.~\eqref{tilde-psi}. Correspondingly,
\be
\dot n_\psi = \frac{\tilde n'_\psi}{a^4} = \frac{m_\psi^2 }{36 \pi t^2}.
\label{n-ferm-phys}
\ee
The energy density of the produced fermions (or antifermions) per unit time is 
\be
\dot \rho_\psi = \frac{M_R \dot n_\psi}{ 2} = \frac{m_\psi^2 M_R}{72 \pi t^2}.
\label{dot-rho-ferm}
\ee
So the width of the scalaron decay into a pair of massive fermions is equal to
\be 
\Gamma_\psi = \frac{\dot \rho^{(\text{tot})}_\psi}{\rho_R} = \frac{2\dot\rho_\psi}{\rho_\Phi} = 
\frac{m_\psi^2 M_R}{36 \pi t^2} \times \frac{6\pi t^2}{M_\text{Pl}^2} = \frac{m_\psi^2 M_R}{6 M_\text{Pl}^2}.
\label{Gamma-psi}
\ee
This result coincides with that of~\cite{gorb-pan} obtained in the Einstein frame, while ours is derived in the Jordan frame.
 
Note, that not only the production {of} $e^+e^-$-pairs by the scalaron oscillations is of interest. There might be much  more efficient production of heavier fermions, including, say, $t$-quarks, as well as  the production of possible heavy sterile neutrinos, $\nu_s$, which are, in particular, popular candidates of warm dark matter particles. The creation of energetic $\nu_s$ with $E \sim M_R/2 \approx 10^{13}$ GeV could make an essential contribution of ultra-high-energy cosmic ray neutrino background.
 

\section{Decays into gauge bosons \label{s-gauge-bosons} }

Under conformal transformation vector gauge bosons are not transformed, $A_\mu \to A_\mu$, and their equations of motion in terms of conformal time are the same as those in flat Minkowski metric. So in this approximation gauge bosons cannot be created by conformally flat gravitational field. This is truth but not all the truth. Conformal anomaly destroys this conclusion and allows for gauge boson to be created~\cite{AD-conf-anom}.

Equation of motion of massless gauge field with an account of the anomaly, as derived in~\cite{AD-conf-anom}, has the form:
\be
A'' - \Delta A - \frac{\alpha \kappa}{\pi} \xi' A' = 0,
\label{A-di-prime}
\ee
where $\alpha$ is the gauge coupling constant squared (for electromagnetic $U(1)$-gauge group $\alpha \approx 1/137$ at low energies),
$\xi = \ln a$, and 
\be
\kappa = \frac{11}{3} N - \frac{2}{3} N_F,
\label{kappa}
\ee 
here $N$ is the rank of the proper gauge group, and $N_F$ is the number of fermion families ($\kappa$ is usually denoted as $\beta$ but here we follow the original paper).

According to the calculations of~\cite{AD-conf-anom} the number density of the produced gauge bosons per unit of physical time is
\be
 \dot n_g = \frac{\alpha^2 \kappa^2}{32 \pi} \left(\frac{\ddot a}{a} - \frac{\dot a^2}{a^2} \right)^2  \approx \frac{\alpha^2 \kappa^2}{32 \pi} \left(\frac{\ddot a}{a}\right)^2,
\label{dot-N-g}
\ee
where we used that $|\ddot a/a| \gg |\dot a^2/a^2|$. Therefore
\be
R = - 6 \left( \frac{\ddot a}{a} + \frac{\dot a^2}{a^2} \right) \approx - \frac{6 \ddot a}{a}.
\label{R-of-a}
\ee 
The last approximate equality is valid for quickly oscillating $R$ given by Eq.~\eqref{R-of-t}.

In~\cite{AD-conf-anom} this equation was applied to particle production near singularity in Friedmann cosmology. Here we shall use it for $R^2$-cosmology. To this end one needs to substitute the average value of 
{$R^2(t)$} taking $\langle \cos^2 (M_R t) \rangle = 1/2$. So
\be
\dot n_g = \frac{\alpha^2 \kappa^2}{32 \pi} \left(\frac{R}{6}\right)^2 = \frac{\alpha^2 \kappa^2}{32 \pi} \times \frac{1}{2} \left(\frac{4 M_R}{6 t}\right)^2 = \frac{\alpha^2 \kappa^2 M_R^2}{144 \pi t^2},
\label{N-g-fin}
\ee
and finally the width of the scalaron decay into two gauge bosons is equal to
\be
\Gamma_g = \frac{\dot \rho^{(\text{tot})}_g}{\rho_R} = \frac{2\dot\rho_g}{\rho_\Phi} = \frac{M_R \dot n_g}{\rho_\Phi} = \frac{\alpha^2 \kappa^2 M_R^3}{144 \pi t^2} \times \frac{6\pi t^2}{M_\text{Pl}^2} = \frac{ \alpha^2 \kappa^2 M_R^3 }{24 M_\text{Pl}^2}.
\label{Gamma-g}
\ee

\section{Conclusion \label{s-concl}}

As it has been shown, see e.g.~\cite{ADR-symmetry}, $R^2$-modified gravity created strong modifications of the universe expansion and cooling laws in comparison with the conventional cosmology governed by the Einstein's General Relativity. In particular, continuous energy influx into the primeval plasma produced by the scalaron decays into great multitude of the different final states resulted in huge dilution of the density of massive stable relics. This phenomenon allows for revival of dark matter (DM) in the form of very massive particles with the typical for supersymmetry interaction strength.

The concrete particle types as bearers of DM are determined by the dominant decay modes of the scalaron. The presented above calculations of the decay widths could help to identify possible types of DM particles for any particular decay channel of the scalaron, if the properties of the final states are fixed by the particle physics model at very high energies.

An interesting feature of the model is a strong production of massive neutrinos, especially of the heaviest unstable ones. They can produce the fluxes of very energetic neutrinos with energies close to the scalaron mass, which migh be observed at IceCube.

\section*{Acknowledgements}
{This work was supported by the Russian Science Foundation under grant no.~22-12-00103.}

\section*{Data availibility statement}
Data sharing is not applicable to this article as no datasets were generated or analyzed during the current study.

\setcounter{section}{1}
\renewcommand{\thesection}{\Alph{section}}
\section*{Appendix  \\ Description of fermions in FLRW space--time \label{ss-ferm}}

\setcounter{equation}{0}
\renewcommand{\theequation}{\thesection.\arabic{equation}}
 
Fermions in curved space--time are usually described in the tetrad formalism which is particularly simple for FLRW space--time. It is the conformally flat metric, which means that after introduction of conformal time~\eqref{conf-trans-1} it is transformed into the form proportional to the flat Minkowski metric~\eqref{FLRW-c}. Under this transformation Dirac equation for massless fermions becomes identical to the equation in flat space--time. So one can conclude that massless fermions cannot be produced by conformally flat gravitational field~\cite{parker1,parker2}. 

In this section we calculate the probability of massive fermion production in FLRW metric following~\cite{GPRT, CB-AD}. The action for fermionic field $\psi$ can be written as
\be 
{ S[\psi] = \int d^4 x\, \sqrt{-g} \, \bar\psi \left( i \Gamma^\mu \nabla_\mu - m_\psi \right)\psi, }
\label{A-psi}
\ee
where {$ \sqrt{-g} = a^3$} is the metric determinant, {$\Gamma^\mu$} is a generalization of the Dirac $\gamma^\mu$ matrices for curved space--time. In the FLRW metric they have the form $\Gamma^0 = \gamma^0$ and {$\Gamma^i = \gamma^i/a$ ($i = 1,2,3 $)}. They satisfy the anticommutation relations $\{\Gamma^\mu, \Gamma^\nu \} = 2 g^{\mu\nu} $, while $\gamma^\mu$ are the usual Dirac matrices, obeying the anticommutation relation $ \{ \gamma^\mu, \gamma^\nu \} = 2 \eta^{\mu\nu} $. $\nabla_\mu$  is the covariant derivative for the spin-1/2 field. In the FLRW metric it is equal to $\nabla_\mu = \partial_\mu + (3/2) (\partial_\mu a)/a$.
 
The conformally transformed fermion field is defined as
\be 
\psi_\text{conf} = a^{3/2} \psi .
\label{tilde-psi}
\ee
In terms of conformal time $\eta$ and field $\psi_\text{conf}$ the action takes the form:
\be 
{ S [\psi_\text{conf}] = \int d^3x\, d\eta\, \bar\psi_\text{conf} \left( i \gamma^\mu \partial_\mu - m_\psi a \right) \psi_\text{conf}. }
\label{A-psi-conf}
\ee
It is the action of a free fermion field in conformal coordinates with mass $m_\text{eff}= m_\psi a (\eta) $.

The time dependence of the scale factor can be found from Eq.~\eqref{H-of-t}, see also Eq.~\eqref{a-of-t-sol}:
\be
a = a_0 \left(\frac{t}{t_0} \right)^{2/3}\,\exp\left[ \frac{2}{3} \int_{t_0}^t \frac{dt'}{t'}\,\sin \left( M_R t' \right)
\right] \approx a_0 \left(\frac{t}{t_0} \right)^{2/3}\,\left[ 1 - \frac{2}{3 M_R t}\,\cos \left( M_R t \right)\right], \quad
\label{a-osc}
\ee
where we have omitted unessential phase $\theta$ and kept the main oscillating term, which is responsible for fermion production.

If we neglect the small and decreasing oscillating term, the scale factor as a function of $t$ evolves as:
\be
a_\text{bg} = a_0 \left( \frac{t}{t_0} \right)^{2/3}.
\label{a-bg-of-t}
\ee
Here the subindex ``bg" means background to distinguish it from total scale factor \eqref{a-osc}. Therefore, 
\be
V = m_\psi a \ \to \ \frac{2 m_\psi a_\text{bg}}{3 M_R t}\,\cos \left( M_R t \right).
\ee

\end{document}